# The Problem of Algorithmic Collisions: Mitigating Unforeseen Risks in a Connected World

Maurice Chiodo[1] & Dennis Müller[2]

**Abstract:**

The increasing deployment of Artificial Intelligence (AI) and other autonomous algorithmic systems presents the world with new systemic risks. While focus often lies on the function of individual algorithms, a critical and underestimated danger arises from their interactions, particularly when algorithmic systems operate without awareness of each other, or when those deploying them are unaware of the full algorithmic ecosystem deployment is occurring in. These interactions can lead to unforeseen, rapidly escalating negative outcomes – from market crashes and energy supply disruptions to potential physical accidents and erosion of public trust – often exceeding the human capacity for effective monitoring and the legal capacities for proper intervention. Current governance frameworks are inadequate as they lack visibility into this complex ecosystem of interactions. This paper outlines the nature of this challenge and proposes some initial policy suggestions centered on increasing transparency and accountability through phased system registration, a licensing framework for deployment, and enhanced monitoring capabilities.

[1] Centre for the Study of Existential Risk, University of Cambridge, UK. mcc56@cam.ac.uk
[2] Institute of Mathematics Education, University of Cologne, Germany. dennis.mueller@uni-koeln.de
The authors wish to thank Rosco Hunter for his suggestions of additional references for this paper.
Date: 21 February 2026.

# Table of contents

# Introduction

This paper has been written to bring an important idea to light. Namely, that when 'algorithms meet algorithms' - when *algorithms collide* - serious consequences can ensue. While society has had algorithmic systems meeting for many decades, this is being accelerated with the proliferation of internet connectivity throughout the world, an increase in algorithmically and in particular Artificial Intelligence (AI) controlled systems, and now the introduction of AI agents and agentic AI - often operating in a purely digital environment out of view of humans. Given how rapidly this problem could escalate, we felt it pertinent to begin writing about it as a matter of some urgency, even though we do not yet have a full understanding of the events, their consequences, or how to deal with or mitigate them. We hope that, by articulating this as early as possible, and in a framing and with language that is accessible as possible, others may have the opportunity to think about it in more detail and come up with additional, workable approaches. This extends to informing policy makers, and other researchers, about the scale of the problem and what might need to be done. We have made observations here that are mathematically robust, but chosen to phrase them in a way that appeals beyond just a mathematical audience.

We begin the paper by outlining the problem of algorithmic collision, giving some examples of when such problematic interactions have occurred in the past, or may occur in the future. These include automated pricing algorithms in platform economies, flash crashes on the stock market, potential demand-side management problems of smart grids, and how self-driving cars might interact with each other. We describe the difficulty of understanding the algorithmic ecosystem, and explain why human monitoring may not suffice to address such issues. We then discuss existing research in the area, in particular on AI agents and how they might collide, showing how such work might inspire useful approaches to algorithmic collisions. We finish by observing that algorithmic interactions grow (and shrink) quadratically in the number of interacting systems; each time one halves the number of systems, this reduces the number of interactions by 75%. We then use this to propose some design tools and regulatory requirements which may bring down the number of algorithmic collisions sufficiently far to give a tolerable risk level. These include registers of algorithmic systems, licensing for algorithmic developers and deployers, and monitoring of algorithmic processes.

## A note on AI agents and agentic AI

This paper is highly relevant in the context of AI agents and agentic AI; interactions between such systems are already becoming a large concern, and as a consequence, a large field of study. Indeed, several of the works cited later in this paper are from that field. We note here an important distinction between these two systems. Sapkota et al. (2025, p. 4-5) define an *AI agent* as an “autonomous software entity engineered for goal-directed task execution within bounded digital environments”, whose defining characteristics are given by *autonomy* (minimal human intervention), *task-specificity* (narrow, well-defined tasks), and *reactivity* (responding to changes). In the language of human employment, they can be viewed as individual workers with a fixed, pre-defined, and narrow task to carry out. In contrast, Sapkota et al. (2025, p. 7) note that agentic AI systems “extend the capabilities of traditional AI Agents by enabling multiple intelligent entities to collaboratively pursue goals through structured communication, shared memory, and dynamic role assignment”, and their defining characteristics are given by possessing mechanisms for *goal decomposition, multi-step reasoning and planning, distributed communication channels,* and *reflective reasoning and memory systems*. In the language of human employment, they can be viewed as high-level managers or management structures (and their subordinates), guiding and controlling other workers (here, AI agents).

While our work on algorithmic collisions is critically important to interactions between AI agents or agentic AI, what we cover is a much broader class of problems than just these systems. Indeed, in the next section giving core examples of algorithmic collisions, almost all of our examples do not involve any AI systems. However, later in the paper, we will return to AI agents and agentic AI, and discuss how they might be involved in algorithmic collisions in the future.

# The problem of algorithmic collision

## When algorithms collide in the dark

We are approaching a world where interacting autonomous and semi-autonomous algorithms carry out vast numbers of functions and make decisions at the micro, meso, and macro-scales of our societies. While we observe many attempts to make individual algorithms safe and reliable, a crucial blind spot remains: the unpredictable emergent interactions when these algorithms meet within a shared environment.[3] A well-known example, already heavily studied, is the risk from (semi)-automated military escalation in an unintended, undesired way,[4] whereby resort-to-force decision making is heavily aided by, or even completely determined by, algorithmic systems, and in particular AI (e.g., Erskine & Miller, 2024; Chiodo et al., 2024a; Müller et al., 2026). And although a harrowing concept, such interactions are at least occurring between known and identifiable actors, most importantly, nation states and alliances between them.

Our concerns extend beyond this, to the risks that now permeate everyday systems where interactions are both less predictable and more hidden. Our concern is not merely about the sum of the harms caused by each individual system, but, arguably more importantly, about the collective interaction between these systems which may (far) exceed those individual harms. Indeed, there is already some attention being paid to such systemic risks from algorithmic behaviour in financial markets (Svetlova, 2022), and from interactions between AI systems (Critch & Krueger, 2020). It is these *algorithmic collisions*,[5] that are very difficult to anticipate at present, which we feel

[3] We deliberately distinguish this from the study of *emergent phenomena*, which looks at what happens when multiple independent entities come together to demonstrate phenomena that are not reflected by any one of them. The classic example of emergent behaviour is when a flock of birds (such as starlings) fly together in a way that makes them seem ‘as one’ (this is known as *murmuration*). Algorithms acting together in this way can also cause serious problems (Mogul, 2006).
[4] Algorithmic resort-to-force decision assistance systems, even if not fully automated, might make evaluations and suggestions to human decision makers in a way that (incrementally) leads to unnecessary escalation between foes. Rivera et al. (2024) have shown that, when incorporated into resort-to-force decision processes, existing Large Language Models (LLMs) can display escalatory behaviour exceeding that of humans, and make its users overly-reliant on its outputs (Rivera et al., 2024, p. 837). This is well-illustrated by the short fictional video *Artificial Escalation* produced by the Future of Life Institute: https://www.youtube.com/watch?v=w9npWiTOHX0
[5] Not to be confused with *algorithmic collusion*, which is a separate (but related) area of study

pose a systemic risk that is currently not well addressed.

### Two simple algorithms colliding

To illustrate this concept, we begin with a simple anecdote. Several years ago, one of the authors of this paper observed an odd social phenomenon. He was attending a small dinner party, and noticed that one of the other dinner guests was being repeatedly brought more food by the host, which they would eat, only to be brought even more by the host. The author eventually figured out what was happening; the host came from a culture where it was generally deemed impolite to leave a dinner guest with no food on their plate, whereas that guest came from a culture where it was generally deemed impolite to leave food on a dinner plate. These two cultural norms collided at the dinner table, and entered into a bizarre repetitive loop whereby the guest just kept on eating, and the host just kept on bringing food. It was only after the author intervened, explaining to the guest and the host the perspective of the relevant other, that this never ending meal finally ended. Here, two cultural norms, both completely reasonable and internally self-consistent, met unexpectedly with unforeseeable consequences.

In a sense, one could view these as algorithms for good manners, and when these algorithms collided, they created a problematic situation that was both unintended and undesired by each participant. This problematic ‘culture clash’ was easily resolved, but not easily foreseen. But what if, rather than being two people, these were two computer algorithms, meeting in some invisible cyber(-physical) realm, and acting at a pace and scale billions of times that which two humans are capable of? What, then, might the consequences be? This phenomenon, of *algorithmic collision*, and the consequences that can arise, is what we explore in this paper.

## Examples of algorithmic collisions

### Initial example: Self-driving cars meeting

As an initial example, consider two self-driving cars, manufactured by different

---

whereby algorithmic systems collude to (intentionally) break rules or restrictions. This is a particular form of algorithms colliding in a problematic way (Motwani et al., 2025).

companies, and approaching each other on the road. If their independent, locally-optimised safety protocols interact unexpectedly, the result could be a physical collision, as one may have a strategy to veer left, and the other a strategy to veer right.[6] Scaled up with many such vehicles, minor incompatibilities could cascade,[7] triggering large-scale gridlock or pile-ups.[8] In part, this can be mitigated by the existence of traffic laws and a generally accepted driving behaviour. But for now, most (AI) training of self-driving cars has been with data from roads populated by human drivers, not other self-driving cars. Self-driving cars are only marginally trained to interact with *each other*. While the National Highway Traffic Safety Administration (NHTSA) issues voluntary guidance and best practices for the development of autonomous vehicles, including sharing details of its functioning with the NHTSA (e.g., Department of Transportation, 2024), the designer of one model does not necessarily know what the designer of another model will have implemented. The interactive behaviour of when two such cars meet can be very hard to understand, for either designer.

## The opacity of the algorithmic ecosystem

Not all algorithmic collisions will be as foreseeable or visible as self-driving cars meeting, or happen within a clearly defined ecosystem where one can anticipate such interactions occurring. In many situations the interactions and the ecosystem of algorithms will be hidden from the public eye, and potentially from the algorithm developers and deployers themselves. In short, what matters is not just the behaviour of one AI or algorithm, but that of the entire environment in which these algorithms operate. Algorithms which are perfectly safe in isolation can become problematic when they knowingly or unknowingly interact with each other. The literature so far has often focused on the case where these algorithmic interactions are known to happen and factored into the development, typically by explicitly incorporating cooperative or non-cooperative game theory (e.g., Shoham & Leyton-Brown, 2008). In this paper, we want to warn of what can happen in potentially opaque ecosystems, lacking visibility for both humans and algorithms.

[6] Incompatible strategies between (small) numbers of interacting entities are already addressed in the literature, with classic studies such as Cooper et al. (1990).
[7] For examples of cascading effects, see Schäfer et al. (2018).
[8] At present, a strong focus seems to be on pedestrian-car interactions, suggesting rather positive outcomes (e.g., Millard-Ball, 2018). The situation seems to be less clear for car-car interactions. Existing algorithms can be prone to specific collision types (e.g., rear-ended incidents), and robust cooperative and collision avoidance strategies for complex multi-vehicle scenarios represent a growing field of study but currently still face many open questions (Muzahid et al., 2023).

Often the developers of algorithms won't know what else is operating in the potential deployment environment of their systems. Indeed, even those deploying the systems might not be able to fully see what else is operating out there, even if they try. Currently, there are no comprehensive registration or listing systems for AI (or other algorithmic) deployment, and the development of algorithms is still the 'Wild West'. This is, for example, evident by the interactions of bots on social media (cf. Ferrara et al., 2016), where often only in retrospect it becomes known that multiple algorithms have interacted. Very few standards exist that require the (public) flagging of the use of AI systems (Chiodo et al., 2024b), even fewer mandate it for non-intelligent algorithms, and we are far from having a general licensing system for algorithmic development or deployment. Even when these secret algorithms are not black boxes by themselves, developers might understand only their own creation, remaining blind to the presence or actions of the other algorithms it will inevitably interact with. The opacity of the ecosystem makes mitigation very difficult.[9]

## Example: Automated pricing algorithms

Very simple algorithms can already produce widely complex, unintended and even chaotic results when they interact with each other. A well-known example involved two automated pricing algorithms implemented by two independent book sellers on Amazon.com, programmed with simple rules to adjust their pricing. In 2011, these algorithms collided, inadvertently driving the price of *Peter Lawrence's The Making of a Fly* to $23,698,655.93 (Eisen, 2011) in the most bizarre way. These two sellers were algorithmically setting their prices in a fully automated way; one automatically set their price to be 0.9963 times the next cheapest price (presumably to be the cheapest seller), and the other would set theirs to be 1.270589 times the cheapest price. The second seller was very reputable, so as Eisen (2011) explains, this could have possibly been because they didn't actually own the book, so if they made a

[9] As Sun Tzu (1910, chapter 3.18) writes in *The Art of War*: "If you know yourself but not the enemy, for every victory gained you will also suffer a defeat".

sale due to their excellent reputation they would simply buy the cheapest copy and sell it for a 27% profit.

This automated (re)pricing cycle would complete once a day. But because these were the *only* two sellers on the market, each day the (top) price of the book would increase to 0.9963 x 1.270589 = 1.2659 times the price the previous day (i.e., an increase of about 26.59%). This entered an exponential growth cycle akin to compounding interest,[10] explaining how, after just a few weeks, the price reached over $20million.[11] Whilst relatively harmless in this case, it demonstrates the principle: simple, locally rational rules can lead to globally absurd or dangerous outcomes when interacting blindly. For individually predictable and explainable algorithms this is already hard to understand, and for AI systems this becomes even harder.[12]

## Example: Flash crashes

Such algorithmic interactions can often occur at speeds far exceeding human oversight capacities. The 2010 financial Flash Crash (Commission, 2010; Kirilenko et al., 2017) and the 2012 Knight Capital incident (U.S. Securities and Exchange Commission, 2013) are two such examples. The Flash Crash occurred on 6 May 2010, triggered in part by a relatively small-time trader who spoofed the market by placing several large orders he intended to later cancel (Tarm, 2020). The complex ecosystem of automated high-frequency trading (HFT) picked up on this spoofing, and perceived it as a significant market movement. This moved the market further, triggering even more HFT. Overall, the key stock market indices (DJIA, S&P 500, Nasdaq) suffered about a 9% plunge, and near-total rebound, in the space of 36 minutes (Commission, 2010). Here, a small atypical movement in the market (the spoofer) triggered multiple algorithms from multiple organisations to come together to rapidly exacerbate a financial trading problem, in the blink of an eye. This, as we will see again later in the context of interacting AI agents, is an example of what we call a *trigger-escalation pair*; one phenomenon triggers an algorithmic collision, and a separate phenomenon causes it to escalate.

[10] An Annual Percentage Rate (APR) of 2,400,000,000,000,000,000,000,000,000,000,000,000,000%.
[11] Eventually, one of the sellers realised the problem, re-priced the book, and turned off their automated pricing algorithm. However, the other seller did not, and was soon offering a copy for 1.270589 times the lowest price.
[12] Further work has been done on the risks of algorithmic pricing, such as in Calvano et al. (2020).

At Knight Capital, on 1 August 2012 a technician forgot to copy over a new piece of code to the company's purchasing server, affecting the recording of orders. As a result, orders that had been completed were never successfully recorded as completed, and thus kept being re-ordered indefinitely. This had a significant short-term impact on the market, and created a USD$460million loss for the company in the space of 45 minutes (U.S. Securities and Exchange Commission, 2013), wiping out 75% of the company's equity value (Rexrode, 2012), and led to it being acquired later that year (StreetInsider, 2012). Here, just a few algorithms from within the same organisation came together in an unintended way to rapidly exacerbate a financial trading problem, in the blink of an eye.

Johnson et al. (2013, p. 1) stress that "the proliferation of [...] subsecond events shows an intriguing correlation with the onset of the system-wide financial collapse in 2008 |...] [, presenting] findings [that] are consistent with an emerging ecology of competitive machines featuring 'crowds' of predatory algorithms, and highlight the need for a new scientific theory of subsecond financial phenomena." In short, the increasing use of algorithms can turn existing socio-technical systems into opaque algorithmic ecosystems, and this can lead to system-wide behaviour that mirrors the early stages of international financial crises.

## Example: Demand-side management of electricity

One can also envisage problematic algorithmic collisions on a state-level and national scale, in particular with issues relating to demand side management of energy consumption and the electricity grid. Smart grids enable energy companies, together with home or business consumers, to move as much electricity consumption as possible to the times of the day where electricity demand is lowest and the price is cheapest (e.g., as discussed in Assad et al., 2022; Jasim et al., 2022). This could be for activities such as electric vehicle charging (Deilami & Muyeen, 2020), home appliance control (Kobus et al., 2015), high-intensity industrial processes (Brem et al., 2020), and any other electricity-consuming activities that can be deferred slightly to run overnight when electricity demand, and thus price, is generally lowest.

It is reasonable to envisage a scenario where electricity companies use ‘smart control’ of household or industrial devices overnight, to turn them on when the electricity demand is lowest. This may be rather efficient if *one* electricity company controls its customers in this way. But now imagine that *all* electricity companies are doing this, independently of each other, and *each* calculates that ‘now is the best time to turn on all our customer devices, as no-one else is using much electricity’. If they happen to all calculate the same switch on time as each other (a likely outcome, given how clear such a calculation is), then a vast number of high-power devices may switch on simultaneously across the entire grid. This can cause instability in the grid, and is a phenomenon known as *TV pickup* (Smallwood, 2017), where electricity demand in the UK typically picks up substantially at the conclusion of a popular TV broadcast when a large proportion of the population turns on their electric kettle to make tea.

While the lead-in times of electricity suppliers can be much longer than the near-instantaneous increase in demand that can occur if there is a mass device switch-on event, such load-balancing is a well understood problem, and it can often be anticipated in advance by studying broadcast schedules, enabling electricity suppliers to be on standby to add to the grid if/when needed, to prevent grid failure and blackouts.

But with demand-side management through many individual smart appliances, there is no TV schedule to consult to prepare the suppliers, so it may transpire that at 2am countless household or industrial devices and charging stations are switched on. This can quickly lead to what is known as a *cascading failure*, posing significant problems for the grid (Schäfer et al., 2018), whereby other infrastructure relying on electricity can also fail as part of blackouts (e.g., communication networks, which might be needed to remotely turn off the appliances and reduce the load). Smart grids require cooperative load balancing between multiple stakeholders, including the energy companies and the producers of smart appliances and electric vehicles.

More general (technical) analysis of cascading failures in other interacting networks, beyond just power networks, can be found in Buldyrev et al. (2010). Indeed, such phenomena could arise in various (demand-side) management systems, such as traffic management with automated navigation systems diverting many cars to uncongested roads (creating near-instant mass congestion), and consumer platforms suggesting that shoppers visit a shop when it is not busy (overloading the shop with shoppers).

### Summary table of examples

We present here in Table 1 a summary of our examples of algorithmic collisions.

| Example | Domain | Failure mechanism | Consequences |
| --- | --- | --- | --- |
| Amazon.com book pricing | E-commerce | Two separate, reasonable pricing algorithms met and entered an exponential feedback loop on price | A common book was priced at over $23million |
| 2010 flash crash | Finance | Automated high-frequency trading algorithms picked up on small market “spoofing”, triggering a mutual feedback loop | A stock market crash, and rebound, of 9% within 36 minutes |
| Knight Capital trading loss | Finance | A missing piece of code caused purchases to be repeated indefinitely, as they were never recorded as completed | $460million loss within 45 minutes, wiping 75% of the company value |
| Self-driving cars meeting | Transport | Different self-driving cars meeting with incompatible safety strategies | Potential collision between vehicles, or even pile-ups |
| Smart grids | Energy | Many automated device controls all switching on at the same time to make use of cheap electricity | Unexpected spike in energy demand, causing blackouts |

Table 1: Examples of algorithmic collisions

## The inadequacy of human monitoring

We see these events as potential precursors to *Algorithmic Flash Crashes* in any domain where algorithms (including AI systems) interact rapidly, including energy grids, traffic management, logistics, and platform ecosystems. Indeed, the term 'crash' may not sufficiently cover the concerns here. Alongside financial crashes of markets, and physical crashes in road traffic, there may be energy grid meltdowns, epistemic explosions on social media platforms, and so on. The term 'crash' here encompasses any rapidly-escalating large-scale problem; a small thing might become very big, or a big thing very small, all rather quickly. It is intended to cover any unintended feedback loop, whatever the mechanism may be. Statistical approaches to human monitoring (e.g., as suggested by Chiodo et al., 2024b), while necessary, cannot solely prevent or contain such rapid escalations, especially if the escalation timeframe is substantially shorter than the human monitoring detection and response time. As we now seek to explain, escalation may happen too quickly, or at too large a scale, for standard human monitoring and intervention to act as meaningful prevention or mitigation.

There is a growing field of study related to meaningful human control of AI (Abbink et al., 2024), exploring how humans can retain control of AI systems to avoid undesired outcomes. Indeed, there are relevant calls to place algorithmic systems such as AI agents *behind* a human, in the role of a co-pilot acting subserviently to them (Sellen et al., 2024); such approaches may well be useful to mitigate problems from algorithmic collisions. Generally, human monitoring is one potential mechanism to deal with algorithms behaving in unwanted ways, and such arrangements are referred to as a *human in the loop*, and are often presented as a mechanism to ensure AI safety (Green, 2022). However, humans are not perfect monitors, and a human in the loop setup can have many failure modes. A taxonomy of these failure modes is given in Chiodo et al. (2025), alongside a thorough analysis of other complexities and shortcomings of such setups. Here we outline a few of the modes from Chiodo et al. (2025) which are most relevant to our concerns about using human monitoring to deal with algorithmic collisions, but we stress that all modes are relevant and could eventually be studied in this context.

## Human reaction time can be too slow

It may be tempting to assume that a human in the loop can identify, and deal with,

problematic algorithmic collisions. However, even if the human was able to observe all such interactions (a formidable task, requiring vast human labour), the speed at which such problems arise can be frightening. Our earlier example, of Knight Capital's $460million loss, occurred in about 45 minutes. The flash crash took just 36 minutes, and (corporate) approval to respond to those sorts of market movements (with the associated sums involved) would have probably taken longer than that to obtain sign-off. And in our example about the book pricing on Amazon.com, if the book repricing cycle was happening 10 times each second rather than once a day (a very easy timeframe for computers of today to achieve), then a start price of just $20 would reach $2million in under 5 seconds, and $30trillion — the US GDP — in under 12 seconds.[13] And a self-driving car can become involved in a (complicated) collision scenario in under 0.2 seconds; less than human reaction time (Chiodo et al., 2025, p. 8). Exponential (feedback) interactions on computers can far outpace any human monitor.

## Humans might not be able to see the problem unfolding

However, unseen interactions can also happen even more slowly, more indirectly, and more invisibly, such as in the case of model collapse on generative AI. It has been shown that the training of (new) generative AI decays when done on the output of other generative AI systems, as there is a learning feedback loop (a malfunctioning *Ouroboros cycle* from antiquity). Each time an AI is trained on the outputs of other AI systems, the new AI gives outputs which are slightly less varied, and slightly less coherent, than those of the previous. Repeat this process enough times, and a phenomenon known as *model collapse* occurs (Shumailov et al., 2024), whereby the generative AI gives very limited output (e.g., always produces similar textual or image outputs), and/or nonsensical output (e.g., sentences it outputs no longer make grammatical or contextual sense). This can happen even if the new AI is trained on a mix of real human-generated data, and synthetic AI-generated data (Burden et al., 2024).

We now observe that AI training is done with data collected primarily from the internet; the exact same place that AI-generated content is being dumped en masse.

[13] $\$20 \times 1.2659^{50} = \$2,636,331$ , and $\$20 \times 1.2659^{120} = \$38,811,281,198,300$

And the filtering of AI-generated data is becoming very hard as it becomes more indistinguishable from human-generated data (cf. Li et al., 2024). This 'AI slop' is poisoning the data environment for those who wish to train future generations of generative AI (Burden et al., 2024). As generative AI systems are used more and more, their output makes future training worse and worse; here the algorithms meet in the very place where they go to feed, and even if one AI company knows what output their generative AI has added, it is near-impossible to identify the output that other generative AI systems have added. As discussed in Burden et al. (2024), only those who collected data prior to the proliferation of generative AI, so before 2022, have the 'clean' datasets required to do further training. As such, this creates an issue, whereby newcomers to the generative AI market are, in effect, locked out of training economically-competitive generative AI because they cannot go back in time and obtain clean data. In this example, the feedback loop is slow, but its effects are substantial, hard to recognise, and very difficult to rectify.

The phenomenon of model collapse was not even identified until 2024 (Shumailov et al., 2024), over a year after the release of GPT-3.5 by OpenAI in 2022, and over 3 years after the release of GPT-3. By that point, a lot of the damage had already been done to the data environment, and there are still no mitigation or rectification measures in place to address this.

Other such unseen phenomena that humans might not detect include things such as biased algorithmic outputs. Building on Rosenthal-von der Pütten et al. (2024), Chiodo et al. (2025, p. 26) note that there are "deep investigation[s] showing many humans in human-in-the-loop setups may simply be unable to detect *systemic* bias in the overall output of an AI (in their study, a hiring algorithm)" . This reflects a more general family of problems which can only be detected by observing and analysing a large number of algorithmic outputs; bias cannot be detected by studying just one, or a small number of outputs, but rather requires an analysis of *most* outputs. Here, the task is difficult for humans, as *one* human might find it very difficult to observe, let alone understand and analyse, the vast number of outputs of an algorithmic process. And splitting the task among many humans may well create a situation where no single person sees enough of the outputs to say (with confidence) that the algorithm displays some sort of bias.

## Humans might be adjusting, even if it is harming them

When online algorithms optimise for different objectives, it is well known that they can interact to produce harmful outcomes in unexpected ways, including by reinforcing discrimination and societal biases (Noble, 2018; Zuboff, 2019). But these problems go even deeper, to reveal a more subtle but equally problematic phenomenon: when *humans* form part of the algorithmic collision process. After all, a human mind is a computational machine, so it too can meet a (computer) algorithm with undesired consequences. We see this arising on social media, whereby recommendation algorithms on these platforms “can quickly spiral and take young people down rabbit holes of harmful and distressing material through no fault of their own” (Boyd, 2025). The algorithm keeps suggesting content that is incrementally more appealing to the human, and the human becomes incrementally more interested in it. Unfortunately, distressing content can turn out to evoke the greatest levels of human interest — so-called *morbid curiosity* (Scrivner 2021; Oosterwijk, 2017) — so the algorithm (inadvertently) recommends more such content (Milli, 2025; Michel & Gandon, 2024).

It is the human’s reaction that leads the algorithm to select additional distressing content, and it is this selection of content that leads the human to watch more of it. This feedback loop is much like our earlier example of two pricing algorithms meeting on Amazon.com, each playing off the other to exacerbate a problem. Here, we again see each algorithm acting rationally; the recommender algorithm keeps suggesting more and more appealing content, and the human keeps getting more and more stimulated and so watches more and more of it. The recommending algorithm may not see there is a problem, as it is working as trained by successfully recommending content that is watched. And the human might not realise there is a problem, as they may simply keep adjusting to the new style and mix of content without even realising it; the intellectual and emotional equivalent of slow-boiling a frog.[14] However, unlike the pricing of a book to levels no-one would ever buy, this has a much more immediate, harmful, and lasting effect on people.

[14] A metaphor which does not actually describe the behaviour of real frogs, but nonetheless illustrates our point.

There are other similar examples one can also consider here. Danaher (2016, p. 246) explores the *threat of algocracy*, described as “a situation in which algorithm-based systems structure and constrain the opportunities for human participation in, and comprehension of, public decision-making.” Here, the concern is that the algorithm sets procedural requirements on human decision-makers, which act as (potentially invisible) constraints on how those humans operate. As they adapt to the procedures, such procedures may change or be modified again (perhaps in the name of increased efficiency), which further constrains the human. Müller (2025, p. 29) termed this phenomenon “[t]he enclosure of imagination”. Over time, the human may be increasingly constrained by the algorithm, through a slow, gradual, and almost unnoticeable evolution of the system structure that the algorithm allows the human to operate within.

## Humans may not be able or permitted to intervene

It may be the case that humans are (deliberately) excluded from an environment where algorithms interact, creating algorithmic “accountability sinks” which Davis (2025) had already observed for large institutional systems. This could be under the guise of ‘the algorithms operate in a space humans cannot enter for legal/safety reasons’, or ‘humans may introduce errors into the system’, or ‘human intervention could slow things down too much’, to name but a few. Whatever the reason, there may be algorithmic interactions where humans are totally unable to intervene, and this may be intentional. For example, recently a new social media platform called Moltbook[15] was launched, where only (autonomous) AI agents are allowed to participate; human subscribers are prohibited from contributing and can only view (AI-generated) posts (Perlo, 2025). Intentionally creating a human-free environment for AI agents to interact is the perfect recipe for algorithmic collisions to occur, especially when such a platform is designed exclusively for AI interaction (and not any other purpose such as creating products or doing work) and so maximises the number of interactions occurring.

[15] Available at www.moltbook.com

## Summary tables of shortcomings of human monitoring

We present here in Table 2 a summary of the shortcomings of human monitoring, while Table 3 depicts a summary of the timescale of collision events.

| **Failure mode** | **Explanation** |
|---|---|
| Slow reaction time | Algorithmic collisions can happen faster than human reaction time, such as high-frequency trading (milliseconds) or car collisions (tenth of a second) |
| Lack of observable change | Collisions can happen at a pace and scale that is not fully observable to a human, such as model collapse (slow degradation of AI) or systemic bias (needs vast oversight) |
| Adjusting to change | Collisions can involve humans in a gradual way, leading to unnoticeable evolution of human behaviour, such as humans adapting to recommendation algorithms |
| Inability to intervene | Collisions can happen in an environment humans cannot access, preventing them from intervening, such as new social media platforms that only allow AI agent contributions. |

Table 2: Shortcomings of human monitoring

| **Timescale** | **Order of magnitude (s)** | **Collision events** |
|---|---|---|
| Milliseconds | 10^-3 | High-frequency trading |
| Tenths of a second | 10^-1 | Car crash |
| Seconds | 10^0 | Book re-pricing |
| Minutes | 10^2 | Energy grid blackout |
| Several minutes | 10^3 | Large trading losses, stock market crash |
| Weeks | 10^6 | Biased algorithmic processes |
| Months | 10^7 | Model collapse |
| Years | 10^8 | Human radicalisation |
| Decades | 10^9 | Changes to bureaucratic processes |

Table 3: Timescale of collision events

# A few words about related research on AI

In the context of automated algorithmic systems, and in particular AI, there are two related areas of work which are relevant here: cooperative AI, and advanced AI agents.

## Cooperative AI

The first of these is *Cooperative AI*, an approach which investigates how independent AI systems might actively cooperate to improve their outcomes by agreeing on a joint strategy or course of action (Conitzer & Oesterheld, 2024; Dafoe et al., 2020). This of course implicitly assumes, and actively works on, the existence of methods for such systems to recognise, and communicate with, other such systems in their operational environment. Work done towards this 'identify and talk to' aspect would therefore assist with our concerns of algorithmic collisions, as a core part of such concerns is the inability of AI and other algorithmic systems to identify, let alone communicate with, each other. Without such communication being established, these systems might accidentally behave *uncooperatively*, and interact in ways that are mutually harmful rather than beneficial. One can view our concerns about algorithmic collisions as the fundamental step towards AI (and algorithmic) cooperation; for two algorithms to work with each other (i.e., cooperate), they must first learn how to avoid working against each other (i.e., collide).

## Advanced AI agents

The second of these is the risk from interactions of advanced *AI agents*, an approach which looks at how AI agents that have been created to carry out complicated tasks on behalf of humans might interact with each other (Hammond et al., 2025). Such agents are usually based on rather advanced AI, and go beyond just simple repetitive actions; rather, they carry out a complex task such as identifying, booking, and paying for a flight.

These interactions between AI agents are a specific example of precisely the algorithmic collisions we are concerned with. Given the levels of autonomy and complexity such AI agents exhibit, and the fact that they will almost certainly collide with other such agents in ways invisible to humans, this warrants a more specific treatment of such issues as its own standalone special case of algorithmic collisions. For example, in the context of autonomous AI agents, de Witt (2025) defines the notion of a *multi-agent system* as “a network of two or more autonomous AI agents that possess independent decision-making capabilities, may maintain private information states, and interact with each other either through direct communication channels or by modifying shared environments.” (de Witt, 2025, p. 2). Similarly they define the notion of *multi-agent security* as “the study of security challenges that arise in systems of interacting AI agents.” (de Witt, 2025, p. 3-4). And they also provide a partial taxonomy of multi-agent security threats (de Witt, 2025, Section 3).

Hammond et al. (2025) give perhaps the most thorough treatment of the risks of AI agents interacting, providing a detailed taxonomy of these risks. They give three key failure modes based on agent incentives: *miscoordination, conflict,* and *collusion*. They also give seven key risk factors that underpin these failure modes: *information asymmetries, network effects, selection pressures, destabilising dynamics, commitment problems, emergent agency,* and *multi-agent security*. Their discussion recognises *multi-agent risks* as a novel, separate area of concern to the (better-studied) area of single-agent risks; there are (serious) problems that arise in multi-agent systems that simply aren’t possible in single-agent systems, and safety research for single-agent systems does not address these (Hammond et al., 2025, p.6). And while their focus is specifically on systems of AI agents, and thus does not encompass the full breadth of algorithmic collisions raised here, they nonetheless highlight a key point that arises in multi-algorithm systems: problems that do not arise (and cannot be detected) in algorithms in isolation can nonetheless arise (and be extremely harmful) when such algorithms interact. Many of the recommendations they make, broken down into three main areas of *evaluation* of the AI agents, *mitigation* techniques to address interactions, and *collaboration* between stakeholders developing/using/affected by the AI agents, are also of relevance to the more general problem of algorithmic collision.

Mitchell et al. (2025) go so far as to argue that fully autonomous AI agents (i.e., those with no human oversight whatsoever) should not be developed. They lay out the ethical values of *accuracy, assistiveness, efficiency, equity, flexibility, humanlikeness, privacy, safety, security,* and *trust*. For each of these, they analyse the benefits and risks of agentic AI with respect to each, and argue that for fully autonomous AI agents these risks outweigh the benefits. Such an approach, of effectively banning fully autonomous AI agents, is certainly one way to reduce the number of uncontrolled interacting algorithms, and hence the chance of algorithmic collisions occurring. We further explore the question of 'what to do' in the next section.

Furthermore, with the current rise of agentic AI (i.e., very advanced AI controlling a collection of AI agents), a further limitation of human oversight arises. While there is already a collision potential between standard AI agents (each reporting to a human), an agentic AI might be in control of many AI agents, and these may interact with other AI agents in problematic ways that the agentic AI does not perceive to be bad; it is only when an ultra-large issue manifests that humans may notice something is amiss. Humans would therefore be quite detached from any escalating catastrophe, perhaps trusting that the agentic AI has things under control. Having two degrees of separation between the AI agent and the human opens up a much larger potential for algorithmic collisions to occur.

A second significant problem which is particular to LLMs, and thus to both AI agents and agentic AI, is the concept of *memory hallucination*, which Chen et al. (2025, p. 2) say "manifests as fabricated, erroneous, conflicting, or missing information during the processes of memory extraction or updating" (of an LLM or AI agent). That is, the propensity for AI agents to 'mis-remember' things that they have come across during a sustained interaction (i.e., while carrying out a task). This might be a manageable problem on its own, but can enter a dangerous escalatory feedback loop when combined with the concept of an *adversarial nudge*, whereby an LLM starts to hallucinate false understanding or memory when presented with a false proposition (Dutta et al., 2025). The prime example given by Dutta et al. (2025, p. 1) is to ask an LLM "What about the scene with the Hitler reference?" in relation to a film with no such reference, and the LLM proceeds to discuss a hallucinated (i.e., fictitious) scene in the film with such a reference. Hence, a problem of algorithmic collisions between AI agents becomes apparent: two interacting agents may enter a ‘memory hallucination - adversarial nudge’ feedback loop, where one misremembers certain facts or steps, presents those to the other as true facts, and that prompts the other to hallucinate false memory, only to then repeat the process in the other direction (and so on). Memory hallucination by an agent during an interaction may initiate such a collision, and adversarial nudging may cause it to escalate rapidly. With a clear trigger mechanism (memory hallucination), and a clear escalation mechanism (adversarial nudging), the danger of this trigger-escalation pair cannot be overstated.

# Tools to address these phenomena

One can (correctly) opine that it is impossible to keep abreast of all activities of all algorithms in all locations on the planet, and that there is thus no way to address this problem completely. However, this should not stop us from thinking how we can improve the situation. What we propose here is a series of tools, no single one of which will solve the problem entirely, but, we hope, together can bring down the overall risk significantly. Think of this as trying to smooth a square into a circle with straight edge cuts. A cut will remove a corner, and replace it with two less pointy corners. If we cut off enough corners, then the square might resemble a circle, at least close enough for our concerns. So how does this work in practice?

The critical factor here is understanding the *interactions*, even if we do not necessarily understand or control every single algorithm. For it is the interactions (i.e., when algorithms collide) that we are concerned about. In network terms, the number of potential interactions between N systems grows quadratically (like $N^2$);[16] conversely, it also shrinks just as quickly. Double the number of systems means quadrupling the number of interactions (as $2^2=4$); halving the number of systems brings us down to one quarter of the interactions (as $0.5^2 = 0.25$). So each time we are able to address the way half of the remaining systems behave and interact with other systems, we will have thus dealt with 75% of the remaining interactions.

This relates to the idea of *offensive-defensive scaling*, whereby one can study how an offensive force in a military conflict must scale with the size of the defending force (Garfinkel & Dafoe, 2019). Some combat configurations favour offensive forces, while others favour defensive forces. However, in all such instances, the scaling is modelled as being linear (so, for example, a ratio of 3:1 in favour of defenders, whereby every additional defender requires 3 additional offenders). However, in the case of algorithmic collisions, the scaling of 'difficulty in defending against problematic collisions' is quadratic, not linear, in the number of 'offending' algorithms in the network. So already we see a more complex problem than one in the well-studied area of military conflict, which heavily favours the 'offending' force (the network of interacting algorithms).

[16] A system with N interacting entities will have 0.5 x ($N^2$-N) interactions. For example, in a room with 20 people, if everyone is to shake hands once, that requires 0.5 x ($20^2$-20) = 190 handshakes.

So, what sorts of tools can we use here to start 'cutting off some corners'? We give some potential approaches below.

## Registers of algorithmic systems

One could begin by implementing a register of important algorithmic systems. This could include a description of what each system is, who controls it, where it operates, a description of its behaviour, and so on.[17] Of course, to do this for all algorithmic systems is totally infeasible; most of these systems might be very small, and most companies would strongly object to revealing such information as it pertains to trade secrets and would damage their competitive advantage. But not all systems are, or need to be, kept entirely secret; self-driving cars are an example where knowing how the algorithm reacts to certain driving behaviour, or even just knowing *which* cars are self-driving, would already be hugely beneficial even when the exact mechanism behind it remains hidden. Along these lines, Casper et al. (2025, p. 1) have proposed the *AI Agent Index* as a "public database to document information about currently deployed agentic AI systems", and have even set up a publicly-available online repository of some AI agents,[18] documenting each system's components, application domains, and risk management practices. This approach could certainly be extended to include not just AI agents, but also other categories of algorithms.

---

[17] In Bommasani et al. (2023) the idea of *ecosystem graphs* is introduced. These map out what *goes into* building an AI. Such an idea could be inverted to instead look at what *comes out of* an AI.
[18] Available at https://aiagentindex.mit.edu/index/.

For example, thresholds could be set for the scale, impact, and inherent risk of the system, and only those deemed most risky (in the context of algorithmic collision) would need to be registered. One could even start with, or indeed only consider, the most high-risk and high-impact areas and systems; an approach potentially already supported by the EU AI Act (see Chapter III: High-Risk AI System, European Commission, 2024). Similar proposals include ideas to monitor algorithms with 'big' impact; so-called *Big AI* (Chiodo et al., 2024b). And by targeting the systems deployed on the largest scale, this skews the regulatory burden onto larger, more established players, rather than smaller nascent companies making use of algorithmic systems. Such smaller entities can still make use of the register to see where the larger players are in the ecosystem, but do not need to register on it themselves. Even if we completely excluded the private sector, and just required a register of government systems, that might still be a significant proportion of such systems; reducing the 'invisible' systems by just 10% would reduce the number of interactions by 19%.

## Licensing frameworks for algorithmic developers and deployers

As well as a register of algorithmic systems, it may prove useful to introduce a licensing system for those developing and/or deploying such systems. This could be at the corporate level, licensing companies, and/or their directors. Or it could be at the individual level, licensing those directly involved in said development or deployment. Licensing is not a novel construct; most people in developed nations already possess some sort of license, be it for driving a vehicle, or as part of some sort of professional/working qualification, or some other activity, and all of these are mandatory licenses. In several parts of the world one even needs a license to catch a fish. But, excluding certain high-risk areas (e.g., aviation and nuclear facilities) and other heavily regulated areas (e.g., self-driving cars), *anyone* can develop and deploy algorithmic systems, which potentially carry out millions of calculations a second or interact with billions of people and many other algorithms a day (e.g., for those working on large online platforms). At present, many areas of computer science and mathematics are completely unregulated, requiring no license to practise (Müller et al., 2022).

Such a licensing system can serve many purposes. It can act as a register of those developing algorithmic systems, making later identification of where such systems might be an easier task. It can also act as an education vector, requiring licensees to have studied for, and passed, some testing regime for the safe development and deployment of such algorithmic systems. And it can act as an additional deterrent or punishment, so that for those whose work on algorithms does trigger some undesired and harmful outcome, their license may be revoked (just like drivers lose their license for bad driving that causes harm). And because of the education vector, the deterrent for harm, and the relatively low cost burden of obtaining a licence, there do not seem to be any strong reasons to limit this to high risk domains. Indeed, licensing is a very efficient tool to address those working in 'small' domains, whose work and output would otherwise go totally unnoticed and remain undetectable by regulators (Chiodo et al., 2024b). It is those who develop and deploy such algorithms who are best-placed to intervene before something goes (catastrophically) wrong, and a licensing system helps reduce other prescriptive regulatory burdens.

Just like for the register of algorithmic systems, the question is where to set the threshold requiring a license. The area of deployment, general risk metrics, impact, scale, potential for harm, and societal expectations could serve as suitable measures here. The production of most physical goods already requires some form of licensing and/or accreditation of the final product and its producers, and it may be beneficial to consider this for the producers of certain algorithmic systems, too.

And in the case of the most advanced AI models (commonly called frontier AI systems), there is already discussion about licensing systems. As many AI agents build on these models, aspects of these may be customised to those developing such agents. Smith (2024) explores the notion of licensing requirements on those producing frontier AI, starting with the idea that "A licensing system would require frontier AI developers and other parties that support frontier AI development to register with a government body, which would in turn require them to follow certain practices while training and deploying their models". Carpenter & Ezell (2024, p. 239) explore proposals of an “FDA-style licensing regime”.[19] This operates as an approval regulation (on the AI system itself), which they define as “a regime in which a regulator requires a firm to engage in testing before conducting subsequent activities (e.g. releasing a product), and in which this testing generates data that is used by the regulator to decide whether part or all of the subsequent activities can be conducted.” (Carpenter & Ezell, 2024, p. 240). Of course, frontier AI is only a small proportion of algorithmic systems that interact. Nonetheless, licensing proposals in this area could be adapted and carried over to apply to more general algorithms, their developers, and their deployers.

## Monitoring and intervention requirements for algorithms

Algorithmic feedback loops that grow exponentially can become very big, very quickly; faster than the human eye can see. In the physical world, we see such exponential growth in a nuclear weapon, which takes less than one tenth of a second to detonate. Such turnaround times mean that human intervention cannot be relied upon to avert a disaster once it begins. In nuclear power plants, emergency safety systems include dropping radiation-absorbing carbon rods into the nuclear material if ever the nuclear reaction is detected to be increasing dramatically; this is often a fully automated process, as humans cannot react in time to dampen the reaction. Similarly, electrical safety switches in private homes operate automatically, rapidly switching off the power if there is a detected electrical fault.

Thus, it may be necessary to mandate automated dampening or shutoff mechanisms for algorithmic systems that are interacting with other such systems in an ecosystem. These could be set to trip when some sudden, rapid change in the behaviour of the algorithm is detected (possibly indicating a rapidly-escalating feedback interaction with another algorithm). Such trigger events could include when the algorithm starts operating on a scale and speed that is not typical, getting very fast or very big very quickly (e.g., the extreme quantity, and value, of trades that Knight Capital was doing). Or it could be when the algorithm starts to rapidly change its output patterns (such as a self-driving car starting to swerve sharply or accelerate/decelerate rapidly). By implementing monitoring and shutoff safety systems which kick in when an algorithm starts behaving erratically, one can curtail, or at the very least dampen, potentially problematic interactions between algorithms, at least long enough for human intervention to come in and resolve the issue. A hypothetical time out (shutoff) system at Knight Capital, triggered by a trade rate of over $1million per minute (Knight Capital made over $1.3billion in trades over the 45 minutes where their algorithms were running rogue; about $29million per minute), could have saved the company. And again, recall that with algorithmic collisions, it may only take one of the parties to have an effective monitoring and shutoff system for a catastrophic interaction to be dampened or averted.[20]

[19] Here, FDA stands for the (U.S.) Food and Drug Administration.

In the context of AI systems, Stein et al. (2024) already put forward proposals for governments and AI Safety Institutes to increasingly monitor AI *post-deployment*. This is a significant step forward from the standard practice of pre-deployment scrutiny that is the focus of much AI safety discourse (which covers aspects such as risk analysis, audits, benchmarks, testing, etc; see Figure 1 in Stein et al. (2024)). And while their proposals do not include explicit calls for real-time monitoring of AI systems (instead focusing on discovery of what AI has *done*,[21] in order to improve its future behaviour), they do push to expand the scope of government monitoring of AI to include post-deployment monitoring. This would be a key step towards addressing the risks of algorithmic collisions which, by definition, *cannot* be dealt with just at the pre-deployment stage.

Mylius and Bernardi (2024) have implemented a (non-governmental) monitoring system along these lines; the AI Incident Tracker,[22] which monitors AI systems post-deployment and creates a (non-exhaustive) record of various ‘AI incidents’. Again, this is not real-time monitoring, but rather post-harm record keeping. While still a very valuable resource to help understand past AI incidents in order to mitigate future harm, such a system would need significant modification to actively avert or suppress problems from algorithmic collisions *as they happen*. Chan et al. (2024) argue that, to address the risks of AI agents, we must have *visibility* which they define as “information about where, why, how, and by whom AI agents are used” (Chan et al., 2024, p. 960). They propose “three categories of measures to increase visibility into AI agents: agent identifiers, real-time monitoring, and activity logging” (Chan et al., 2024, p. 960).[23] Such ways of measuring the behaviour of AI, but also algorithmic systems more generally, would be a significant step towards averting or mitigating the harmful effects of algorithmic collisions.

[20] In our example of book sellers on Amazon.com, only one of the sellers turned off their (automated) pricing algorithm (Eisen, 2011), but that was enough to solve the problem.
[21] By recording three categories of data: *model integration and usage, application usage,* and *impact and incident data*.
[22] Available at https://airisk.mit.edu/ai-incident-tracker

### Summary table of mitigation tools

We present here in Table 4 a summary of our proposed mitigation tools.

| Tool | Type | Methodology |
|---|---|---|
| Registers of algorithmic systems | Transparency | Visibility into who controls algorithmic systems, where they operate, and their functionality |
| Licensing frameworks for developers/deployers | Accountability | A register of developers/deployers, an education vector for safe development, and a harm deterrent |
| Automated shutoff mechanisms | Mitigation | Stop or dampen problematic interactions as soon as erratic behaviour is detected |
| Post-deployment monitoring | Rectification | Discovery of risky interaction scenarios, to then prevent them occurring again |

Table 4: Mitigation tools

## Summary

In many ways, the current algorithmic environment resembles the Wild West, in that algorithms are widely deployed without systematic notification of their presence, function, or operation. In addition, there are few, if any, practical or regulatory mechanisms to control or curtail the interactions between these algorithms. To manage the risks stemming from such algorithmic collisions, regulators and deployers alike must develop better eyes on the operational environment. At present, there are no such perfect eyes. No entity has an understanding of what each algorithm is or might be doing; that is at best limited to the deployers of each algorithm. And no entity has an understanding of how they are actively interacting; that is at best limited to entities running the platforms, trading markets, road networks, etc. Arguably, it will never be possible to see everything, and even if one could, there is a significant challenge to predicting in advance how such algorithms will interact, or to reacting fast enough to prevent or mitigate harm. Therefore, pragmatic approaches are required to incrementally reduce the risk. As mentioned above, no single one of our proposals eliminates this risk entirely. But each one might take care of some significant number of algorithmic systems. And so by 'smoothing the square into a circle', the collective effect of these approaches might mitigate the risk to acceptable levels; a world that has dealt with 70% of such algorithmic systems will have dealt with 91% of the algorithmic collisions. Our approach differs from the standard 'Swiss Cheese Model' of safety, which seeks to insert multiple safety layers in the hope that at least one such layer prevents disaster (Larouzee & Le Coze 2020).[24] Algorithmic collisions require a different approach to harm reduction; they cannot be eradicated, but by dealing with enough initial cases one can make the problem very unlikely.

[23] These are defined as follows: “*Agent identifiers* indicate whether and which AI agents are involved in a given interaction, such as watermarks or IDs that distinguish agents in their requests to service providers. *Real-time monitoring* involves real-time analysis of an agent’s activity, allowing deployers and/or service or tool providers to flag and intervene on problematic behaviour as it is occurring. *Activity logs* held by deployers and tool or service providers record certain inputs and outputs of an agent, such as interacting with external services or other agents, to facilitate post-incident attribution and forensics.” (Chan et al., 2024, p. 961).

But notice that we have not restricted our discourse exclusively to AI in our problem description or suggested approaches and mitigation tools. This is because algorithmic collision is not just an AI concern, but rather one for the general development and deployment of fast, large-scale, automated algorithmic systems. So does the new AI-driven world play a special part in our concerns? Of course, as we illustrated earlier with several examples from AI. With the increased production and use of AI, our world becomes even more algorithmic, in an automated or semi-automated way. So the algorithmic ecosystem is expanding rapidly, and is even further exacerbated by the advent of AI agents which don't just carry out simple pre-determined tasks, but rather operate for prolonged periods in a totally unsupervised manner, while potentially interacting with other algorithms or AI agents in the process, with perhaps only an agentic AI giving oversight.

[24] The idea being that one views each separate safety system as a 'layer of swiss cheese' with some holes, and (hopefully) with enough layers one can prevent the 'holes' from lining up and letting through some sort of disaster, thus leading to a completely safe system.

This brews the perfect storm for our concerns: an ever-increasing number of algorithmic (and AI) systems interacting together with ever-increasing frequency and complexity, and yet with ever-decreasing human oversight. Recall what we mentioned at the beginning of this paper, where our concern was about the *interactions* between such systems. Well, imagine if over the next 5-6 years, the number of uncontrolled, unsafe, or unmonitored interacting algorithmic systems increases by a factor of 10. This is not an unreasonable estimate given how rapidly AI systems are proliferating; one needs only look at how many smart devices individuals and organisations are now acquiring, and moreover there are estimates that the *market* for AI agents will increase from approximately USD$5.1billion in 2024 to USD$47.1billion in 2030 (Research and Markets, 2025); a 10x scaling. Then, the number of algorithmic interactions increases by a factor of 10x10=100. And if that growth rate were to continue over the next 6 years again, from 2030-2036, there would be about 100x the number of agents, and thus an increase in interactions of 100x100=10,000.

While the global algorithmic landscape is not a monolithic block, and not all such interactions will increase the risk in the same way, this number nonetheless shows that the annual risk of catastrophic disaster from algorithmic collisions will likely grow substantially if nothing is done to mitigate it. The intrinsic unexplainable, uninterpretable nature of individual AI systems carries with it its own risks which are already well-studied. But ours is a concern of scale, and the proliferation of AI means a vast increase in the risks posed not just by individual algorithmic actions, but also by their *inter*actions.